\documentclass{elsart}

 \usepackage{epsfig}

\usepackage{amssymb}
\usepackage{lscape}

\begin{document}

\begin{frontmatter}



\title{The $^3$He($\alpha$,$\gamma$)$^7$Be S-factor at solar energies: the prompt $\gamma$ experiment at LUNA}


\author[ge]{H. Costantini\corauthref{cor}\thanksref{add}}
\thanks[add]{Dipartimento di Fisica, Via Dodecaneso 33, 16146 Genova, Italy,}
\corauth[cor]{Corresponding author Address:  Telephone:+39 010 353 6336; fax: +39 010 314218.}
\ead{costant@ge.infn.it}
\author[dresden]{D. Bemmerer}
\author[ge]{F. Confortola}
\author[lngs]{A. Formicola}
\author[atomki]{Gy. Gy\"urky}
\author[lnl]{P. Bezzon}
\author[mi]{R. Bonetti\thanksref{now}}
\thanks[now]{Deceased}
\author[pd]{C. Broggini}
\author[ge]{P. Corvisiero}
\author[atomki]{Z. Elekes}
\author[atomki]{Zs. F\"ul\"op}
\author[to]{G. Gervino}
\author[mi]{A. Guglielmetti}
\author[lngs]{C. Gustavino}
\author[na]{G. Imbriani}
\author[lngs]{M. Junker}
\author[lngs]{M. Laubenstein}
\author[ge]{A. Lemut}
\author[na]{B. Limata}
\author[pd]{V. Lozza}
\author[mi]{M. Marta}
\author[pd]{R. Menegazzo}
\author[ge]{P. Prati}
\author[na]{V. Roca}
\author[bo]{C. Rolfs}
\author[pd]{C. Rossi Alvarez}
\author[atomki]{E. Somorjai}
\author[te]{O. Straniero}
\author[bo]{F. Strieder}
\author[ca]{F. Terrasi}
\author[bo]{H.P. Trautvetter}

\address[ge]{INFN and Dipartimento di Fisica Universit\`a di Genova, Genova Italy}
\address[dresden]{Forschungszentrum Dresden-Rossendorf,
Dresden, Germany}
\address[lngs]{INFN, Laboratori Nazionali del Gran
Sasso, L'Aquila, Italy}
 \address[atomki]{ATOMKI, Debrecen, Hungary}
\address[lnl]{INFN, Laboratori Nazionali di Legnaro, Padova, Italy}
\address[mi]{Istituto di Fisica Generale Applicata, Universit\`a di
Milano \& INFN Milano, Milano, Italy}
\address[pd]{INFN Padova, Italy}
\address[to]{Dipartimento di Fisica Sperimentale, Universit\`a di
Torino \& INFN Torino, Torino, Italy}
\address[na]{Dipartimento di Scienze Fisiche, Universit\`a ``Federico II'' \& INFN Napoli,
Napoli, Italy}
\address[bo]{Institut f\"ur Experimentalphysik III, Ruhr-Universit\"at
Bochum, Bochum, Germany}
\address[te]{INAF, Osservatorio
Astronomico di Collurania, Teramo, Italy}
\address[ca]{Dipartimento di Scienze Ambientali, Seconda Universit\`a
di Napoli, Caserta \& INFN Napoli, Napoli, Italy}

\begin{abstract}

The $^3$He($\alpha$,$\gamma$)$^7$Be process is a key reaction in both Big-Bang nucleosynthesis and p-p chain of Hydrogen Burning in Stars.
A new measurement of the $^3$He($\alpha$,$\gamma$)$^7$Be cross section has been performed at the INFN Gran Sasso underground laboratory
by both the activation and the prompt $\gamma$ detection methods. The present work reports full details of the prompt $\gamma$ detection experiment, focusing on the determination of the systematic uncertainty. The final data, including activation measurements at LUNA, are compared with
the results of the last generation experiments and two different theoretical models are used to obtain the S-factor at solar energies.

\end{abstract}

\end{frontmatter}


\section{Introduction}

The $^3$He($\alpha$,$\gamma$)$^7$Be reaction is the onset of the $^7$Be and $^8$B branches of the pp-chain in Hydrogen burning
from which the $^7$Be and $^8$B neutrinos are generated. Thanks to the recent precise measurements performed by SNO and
 SuperKamiokande \cite{harmin,hosaka}, the $^8$B neutrino flux is known with a 3.5\% of uncertainty, while the $^7$Be neutrino flux will be measured by Borexino and Kamland in a near future with similar precision \cite{borexino,kamland}. The solar neutrino flux depends on both astrophysical inputs,
such as the luminosity, the radiative opacity, the diffusion and the elemental composition, and on nuclear physics inputs, i.e. the rates of
 nuclear reactions involved in the pp-chain. The
uncertainty on the input parameters directly translates into uncertainties in the neutrino
flux prediction. To obtain information on the astrophysical parameters from the solar neutrino flux, it is therefore necessary to know
the nuclear reaction rates with an uncertainty similar to that of the measured neutrino flux.

Furthermore the $^3$He($\alpha$,$\gamma$)$^7$Be is a fundamental reaction in Big-Bang Nucleosynthesis (BBN), since, according to the Standard Model of BBN, $^7$Li is produced almost exclusively by the $^3$He($\alpha$,$\gamma$)$^7$Be
reaction followed by the decay of $^7$Be. 
The large discrepancy of more than a factor two between the predicted and the observed $^7$Li abundance \cite{spergel} is up to now, not understood. While it is unlikely that the explanation could come from a better knowledge of the $^3$He($\alpha$,$\gamma$)$^7$Be reaction rate, the latter represents the necessary basis of the serch for possible different solutions to the $^7$Li problem.

The  $^3$He($\alpha$,$\gamma$)$^7$Be  reaction is a capture process that occurs through the formation of a $^7$Be nucleus with the
emission of $\gamma$-radiation coming from the direct capture into the ground state and into the first excited state of $^7$Be.
The $^7$Be decays by EC to the first excited state of $^7$Li with a branching ratio of 10.44 $\pm$ 0.04\%  \cite{tilley} and subsequently emits a $\gamma$ of 478 keV.
In the last forty years, the reaction has been studied either detecting the prompt $\gamma$ rays or detecting the delayed $\gamma$ from
the decay of $^7$Be. The overall analysis presented in \cite{adelberger} quotes an uncertainty on the $^3$He($\alpha$,$\gamma$)$^7$Be
reaction rate coming from the discrepancy between the results obtained by measuring the reaction using the above two methods.
This uncertainty (9\%) has been the highest among the nuclear physics inputs adopted in the SSM \cite{bahcall}.

In the last four years a new series of measurements has begun, starting with an activation measurement \cite{nara}. These new
studies tried to measure the reaction with high precision and therefore to investigate the possible discrepancy between the two
 techniques that could be given either to some underestimated  systematic errors or to some possible non radiative transitions \cite{E0,noE0}.
 The aim of our experiment was therefore to provide high precision data obtained simultaneously using both methods.
Here we present with full details the prompt $\gamma$ approach focusing on the analysis of systematic errors.

\section{The experimental setup}
\label{1}

\begin{figure}
    \includegraphics[angle=-90,width=1.1\textwidth]{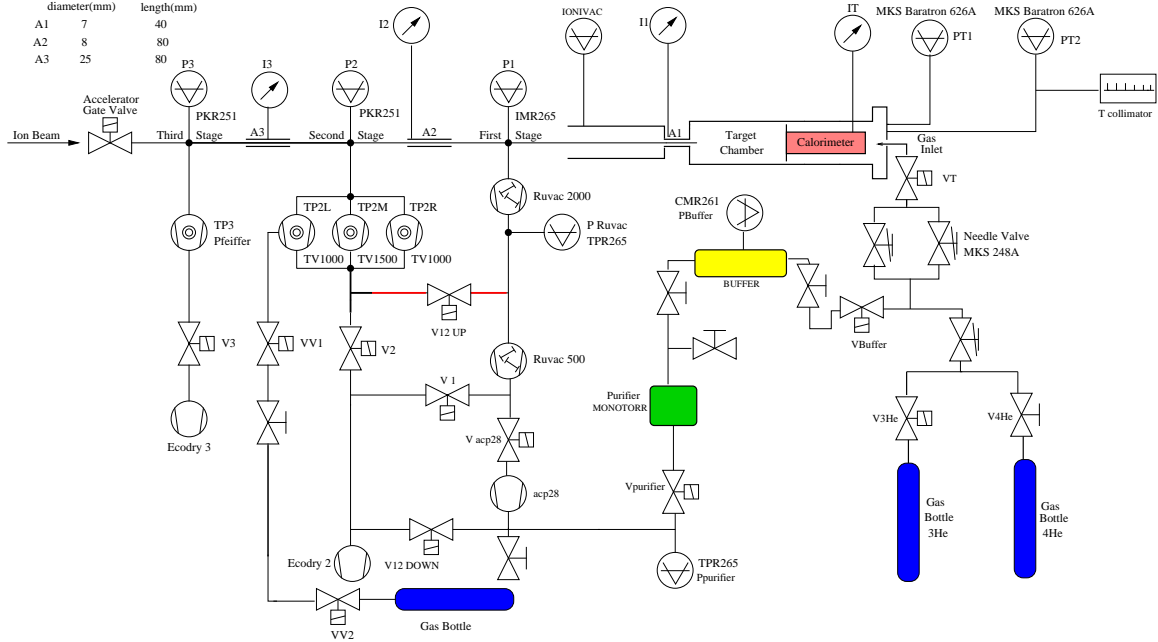}
    \caption{Schematic drawing of the setup with windowless gas target including the three pumping
stages, the interaction chamber and the $^3$He recirculation and
purification system  \cite{tesi_fede}.
  }
    \label{schema}
\end{figure}

The simultaneous measurement of the prompt and the delayed $\gamma$ of
the $^3$He($\alpha$,$\gamma$)$^7$Be reaction, was carried out at
the Laboratory for Underground Nuclear Astrophysics (LUNA) situated deep underground at the
Gran Sasso INFN Laboratory (LNGS). The unique cosmic background suppression offered by 3800 meters water equivalent
rocks of the Gran Sasso mountain, has given the possibility in the last two decades to
measure several nuclear reactions  belonging to the pp chain and CNO cycle of Hydrogen burning in stars \cite{bonetti,casella,formicola,lemut}.

The $^3$He($\alpha$,$\gamma$)$^7$Be reaction cross section was
studied at energies E$_{\alpha}$=220,
250 and 400 keV using the 400 kV LUNA2 accelerator which delivers an $\alpha$ beam of
approximately 250 $\mu$A with an uncertainty on the energy calibration of 300 eV
\cite{acceleratore}.
The measurement was performed using an extended windowless $^3$He
gas target setup. The gas target system has already been described
elsewhere\cite{giuri}. Briefly, it consists in a series of
differential pumping stages separated by high flow impedance collimators (A1, A2 and A3 in figure \ref{schema}) that allow the pressure to drop from
a typical value of 0.7 mbar in the target chamber to 10$^{-6}$ mbar, that is the pressure of the accelerator tube.
During the experiment the $^3$He gas was recovered from the first
and the second pumping stages, purified through an industrial
purifier (Saes Getter MonoTorr II), and fed back to the target chamber (see figure \ref{schema}).
The pressure inside the target chamber was continuously monitored
during the experiment with capacitance gauges at two different
positions (PT1 and PT2 in figure \ref{schema}): one close to the entrance collimator, and an other approximately at
the center of the target chamber. The pressure and temperature profile inside the target
chamber and in the connecting pipe between the interaction chamber
and the first pumping stage have been measured with a dedicated
chamber identical to the one used during the measurements, but
with several apertures along the target length. From these
measurements the target thickness without beam was obtained with
an uncertainty of 0.8\% \cite{giuri}. Due to the intense $\alpha$
beam, the target density along the beam path was decreasing due to
the well-known beam heating effect \cite{goerres}. This phenomenon
was investigated using a 100 $\mu$m  thick silicon detector positioned
inside the target chamber, detecting the projectiles elastically
scattered first in the target gas and subsequently in a movable 15
$\mu$g/cm$^2$ carbon foil. This effect was measured at different
target gas pressures and at different positions in the target along the beam path.
Details on the elastic scattering measurements are described
elsewhere \cite{marta}. The purity of the target was also
monitored  using the elastic scattering \cite{marta} and during the whole
experiment the nitrogen contamination always remained below 2.7\%.
The overall uncertainty on the target
density considering the without-, the with-beam density
measurements and the uncertainty on the gas purity corrections, is of 1.5\%.

The beam entered the interaction chamber through a 7 mm diameter
collimator and was stopped on a detachable copper disk that served
as the primary catcher for the produced $^7$Be and as the hot side
of a calorimeter (see figure \ref{fig:camera}). The latter was used to measure the beam
intensity from the difference between the calorimeter power values
with and without beam and was similar to the one previously used
\cite{lemut}. The calorimeter was calibrated in the whole energy
range, using the evacuated target chamber as a Faraday cup. The
calibration was periodically repeated during the entire
measurement. The reproducibility of the calibrations was within
1.5\%: this value was adopted as the uncertainty on the beam current
determination.

\section{The background reduction}

\begin{figure}
    \includegraphics[width=\textwidth]{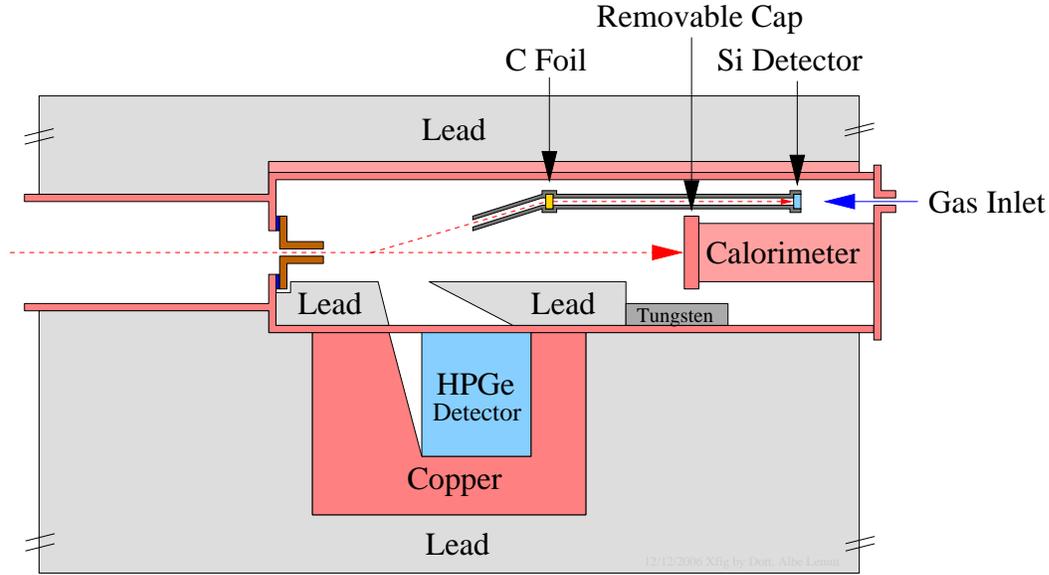}
    \caption{Schematic view of the interaction chamber with the position
           of the HPGe detector and of the 100 $\mu$m silicon detector used for $^3$He
   	 density monitoring. The distance between the entrance  collimator and the calorimeter is 35 cm.
  }
    \label{fig:camera}
\end{figure}

\begin{figure}
    \includegraphics[width=\textwidth]{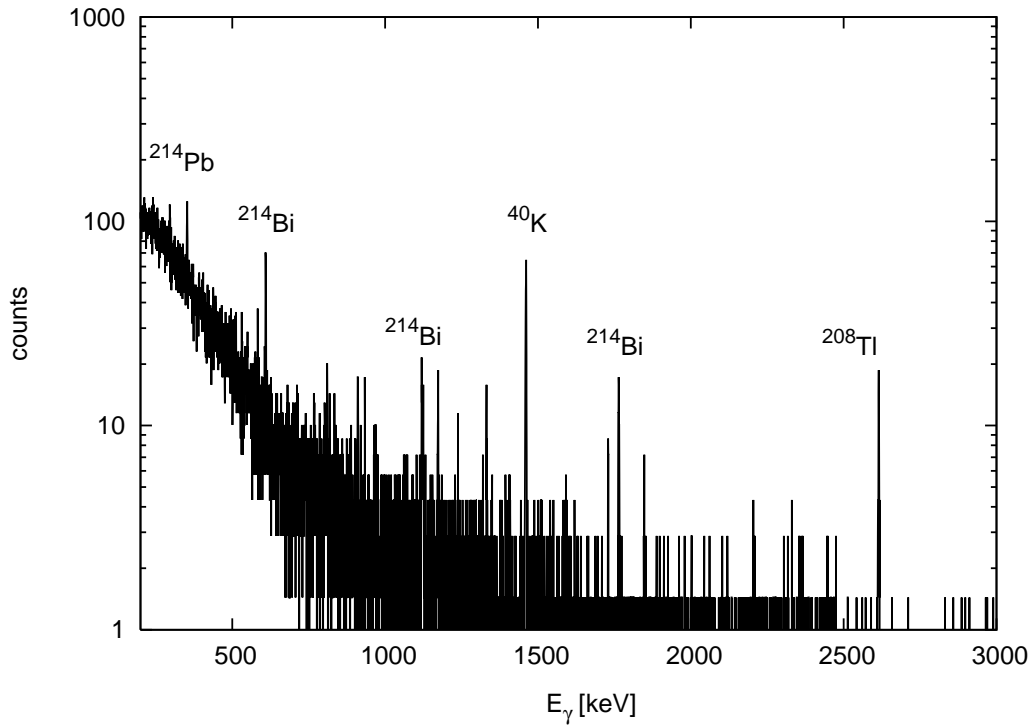}
    \caption{Laboratory background spectrum taken with the shielded 137\% HPGe detector. Measuring time was 31 days and counting rate
 was 0.05 counts/day/keV and 0.11 counts/day/keV in the energy region of interest for the transition to
the ground and to the first excited state, respectively. At the lowest explored energy (E$_{\alpha}$=220 keV) the reaction rate was 1.04 counts/day/keV and 0.57 counts/day/keV for the transition to the ground and to the first excited state, respectively.}
    \label{spettro}
\end{figure}

The prompt $\gamma$ rays coming from the direct capture to the first
excited state and ground state of the $^7$Be nucleus, were
detected by a 137\% (relative efficiency) HPGe detector (figure \ref{fig:camera}) positioned with its front face 7 cm from the beam axis.
 Since the energies of the prompt $\gamma$ rays (0.4, 1.3 and 1.7 MeV) are in the energy region of natural radioactive
isotopes, a massive 0.3 m$^3$ copper and lead shielding was built around the
detector and the target chamber. Passive shielding is particularly
effective underground since the muon flux, coming from cosmic
rays that, at surface, produces energetic neutrons which, in turn, may give rise to $\gamma$ rays in the
lead, is reduced by six orders of magnitude in the Gran
Sasso laboratory. The entire shielding was enclosed in a anti-radon
envelope, which is a plexiglas box flushed with N$_2$ gas to avoid
$^{222}$Rn background. Similar shielding was used for the off-line measurements (activation method). To further reduce the background on the
detector, the target chamber was built with oxygen free high conductivity copper (OFHC) and
no welding materials were used in the chamber assembly. Moreover
low activity materials were used to build the silicon detector
support and all the equipment inside the target chamber (figure \ref{fig:camera}).In this way, a background suppression of 5 orders of magnitude
 was reached for $\gamma$ rays below 2 MeV with respect to a background spectrum measured underground with no shielding \cite{caciolli}. Figure \ref{spettro} shows the background spectrum.
 Aside from radioactive isotopes, background could come also from
 beam induced reactions. A background measurement at E$_{\alpha}$=400 keV
 substituting $^3$He gas with inert $^4$He gas was performed but no additional counts were detected with respect to the laboratory background.
Further details regarding the $\gamma$ ray background can be
found elsewhere \cite{caciolli}.

\section{Angular distribution effects and detection efficiency}

According to DC model calculations \cite{parker-teo} the
$^3$He($\alpha$,$\gamma$)$^7$Be direct capture mainly proceeds by
E1 transition that can occur through s- or d-waves. The angular
distribution function W($\theta$) can be expressed as:

\begin{equation}
W(\theta)=1+a_1P_1(\theta)+a_2P_2(\theta)+\cdots
\end{equation}

where a$_1$ and a$_2$ are the coefficients of the Legendre
Polynomials P$_1$($\theta$) and P$_2$($\theta$).
  To minimize the systematic error due to
angular distribution uncertainty,
a lead collimator has been inserted inside the target chamber
(figure \ref{fig:camera}) to collect at the HPGe detector mostly the $\gamma$
 rays emitted around 55$^\circ$, angle at which the second Legendre Polynomial vanishes.

\begin{figure}
\begin{center}
    \includegraphics[width=1.\textwidth]{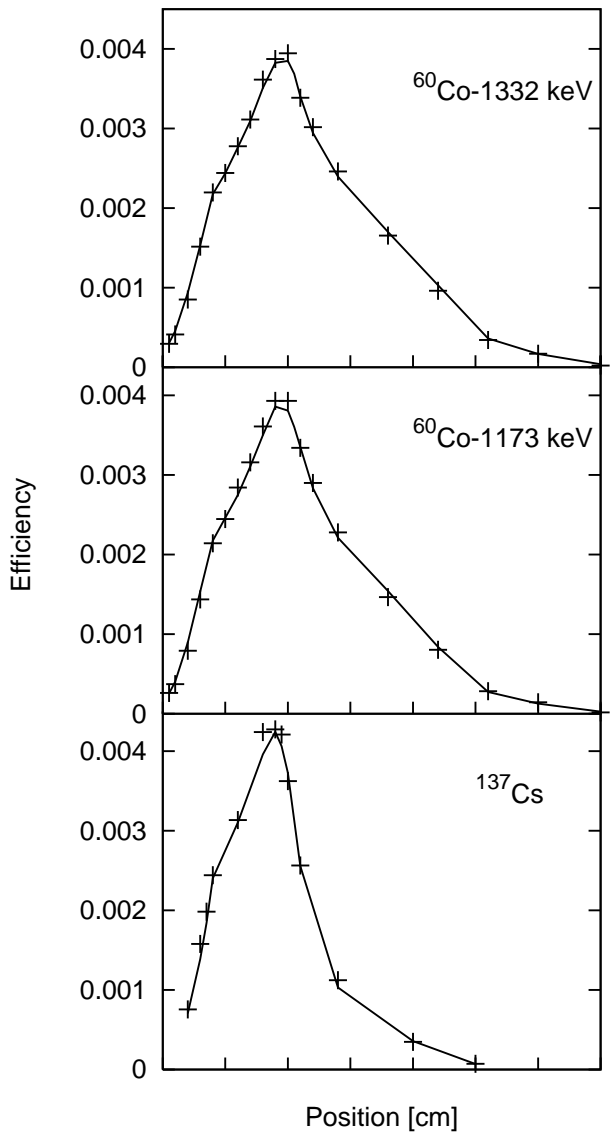}
    \caption{Efficiency profiles measured using point-like sources of $^{60}$Co and $^{137}$Cs with the inner lead and tungsten collimators in the chamber.
 Crosses represent the experimental data while lines are
linear interpolations of MC calculations. The zero position corresponds to the entrance of the beam inside the target chamber.}
    \label{eta}
\end{center}
\end{figure}

\begin{figure}
    \includegraphics[width=\textwidth]{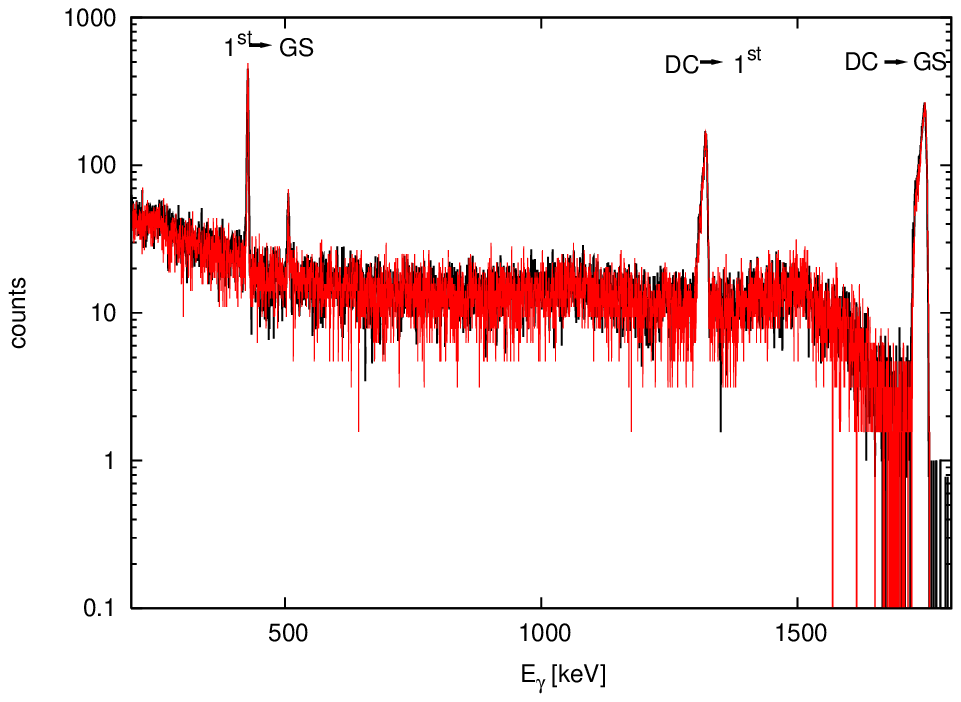}
    \caption{Comparison between the experimental (black curve) and the simulated (red curve) $\gamma$ spectrum at
E$_{\alpha}$= 400 keV and P$_{target}$=0.7 mbar. The simulated spectrum was normalized to the experimental one at E$_{\gamma}$=1.76 MeV to allow the shape comparison.}
    \label{spettri}
\end{figure}

This collimator is a lead brick, 3 cm thick, with a hole shaped as a
truncated cone with elliptical base and the main axis inclined
with respect to the vertical of 45$^\circ$ (figure \ref{fig:camera}). This particular shape was
studied with the LUNA Monte Carlo (MC) code \cite{MC} taking into
account the extended target effect and the detector solid angle,
which depends on the HPGe detector dimensions and its distance from the
beam. The lead collimator and a tungsten brick (1.6 cm thick) were
positioned in the target chamber also to shield the detector from
possible beam induced radiations coming from the calorimeter cap,
and from laboratory background coming from the upstream and
downstream apertures in the shielding.
 In an extended gas target, the interactions are taking place
along the whole beam path and each interaction has a different
geometrical subtending angle to the detector. The detection
efficiency profile $\eta$(z) has been measured moving a $^{60}$Co (E$_{\gamma}$= 1173, 1332 keV)
and $^{137}$Cs (E$_{\gamma}$= 662 keV) point-like sources along the
beam axis from the collimator to the calorimeter cap.
Due to the particular shape of the inner lead
collimator, the efficiency profile along the target length was
quite complicated and the LUNA Monte Carlo simulation code was used to evaluate the detection efficiency for
the $^3$He($\alpha$,$\gamma$)$^7$Be $\gamma$ lines. The
crucial point in the simulation has been the HPGe description and
in particular the determination of the active volume of the
detector, information not provided by the
manufacturer. To determine this parameter, the inner
collimator was removed from the chamber and a first set of
efficiency measurements was performed using the calibrated point-like
sources placed in several points along the beam path. By comparing the MC simulations with the results of these first measurements, the detector
geometry was determined. Subsequently, measurements and simulations were performed with
 the inner lead collimator. A comparison between the simulated
and the experimental efficiency profiles $\eta$(z) is shown in
figure \ref{eta}.
 In the data analysis the integrated efficiency profile along the target length L was used (see eq.\ref{analysis}). The percentage difference between the simulated and experimental integrated efficiency profiles is defined as:

\begin{equation}
\Delta_{int}=\frac{\int_0^L{\eta_{sim}(z)dz} - \int_0^L{\eta_{ex}(z)dz}}{\int_0^L{\eta_{ex}(z)dz}}.
\end{equation}

and it turned out to be (0.3$\pm$1.5)\% and (0.6$\pm$1.5)\% for the 1173 and 1332 keV $\gamma$ lines of the $^{60}$Co source, respectively and (-0.4$\pm$1.5)\% for the 662 keV line of the $^{137}$Cs source.
 The simulation reproduced the integrated experimental
efficiency within the source activity uncertainties (1.5\%).
With the detector geometry fixed through the comparison with the $^{60}$Co and $^{137}$Cs sources, and the detailed description of the target geometry
(i.e. inner Pb and W collimator geometry), the simulation reproduced the experimental $^3$He($\alpha$,$\gamma$)$^7$Be $\gamma$ spectra
at the level shown in figure \ref{spettri}.
Summing effects between the primary and the secondary $\gamma$ transitions in the DC$\to$429$\to$0 cascade, actually smaller than 1\%, were considered in the MC simulation and included in the data analysis.

 Angular distribution functions have previously been calculated
down to 210 keV \cite{parker-teo} and showed a small anisotropy
for both the transition to the first excited state ($\gamma_1$)
and to the ground state ($\gamma_0$).  Experimental measurements
carried out down to E$_{cm}$=148 keV \cite{krawinkel} confirmed
the anisotropy manifesting interference effects of both partial
wave contributions. Recent theoretical predictions \cite{kim} are
in agreement with the theoretical angular distribution functions
of \cite{parker-teo}. Predictions of a$_1$ and a$_2$ can be found in \cite{parker-teo}
as a function of the incident beam energy. These curves
 have been linearly extrapolated down to 200 keV
and the coefficients of the Legendre polynomials adopted in the
detection efficiency calculation are a$_1$ = -0.05 and a$_1$ = 0
for the transition to the ground and to the first excited state,
respectively, and a$_2$ = -0.1 for both transitions. To estimate
the effect on the detection efficiency of the uncertainty on the
angular distribution, we have varied both a$_1$ and a$_2$ coefficients in the Monte Carlo simulation and
100\% changes resulted in a global 2.5\%
variation of the detection efficiency. The latter has been assumed as a systematic uncertainty and turned out to be the major
contribution to the error budget of the prompt $\gamma$ experiment.

\begin{figure}
    \includegraphics[width=\textwidth]{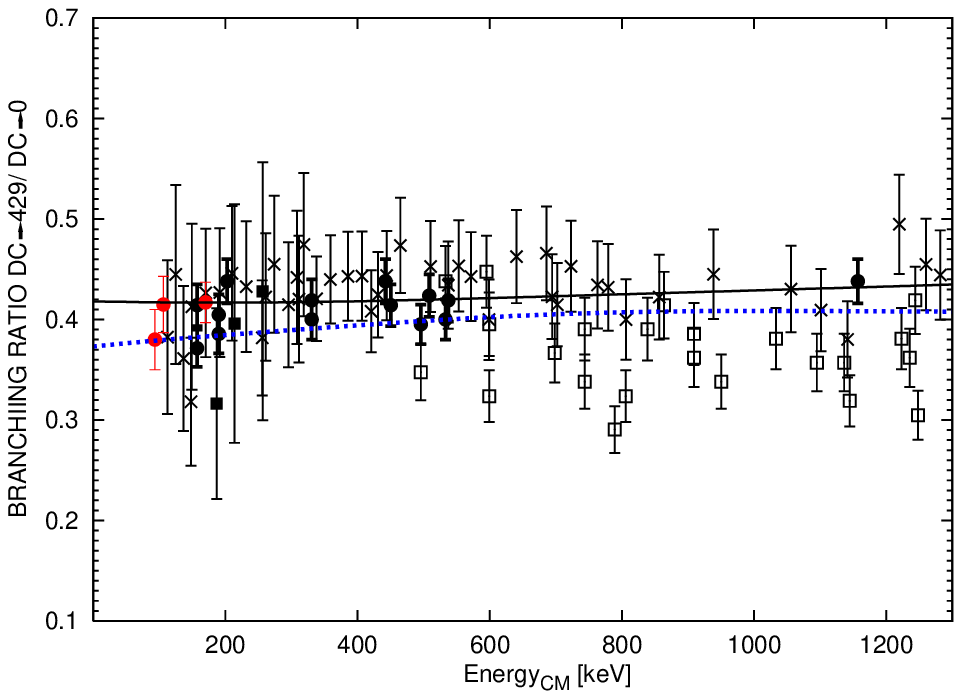}
    \caption{Branching ratios for the $^3$He($\alpha$,$\gamma$)$^7$Be reaction. The present data (red dots) are compared to previous experimental results from  Parker and Kavanagh \cite{parker-exp} (open squares), Nagatani et al. \cite{nagatani} (filled squares), Kr\"awinkel et al. \cite{krawinkel} (crosses) and Osborne et al. \cite{osborne} (black dotes). The solid and dotted curves are from the calculations of Liu et al. \cite{liu} and of Kajino et al.\cite{kajino1} respectively. }
    \label{branching}
\end{figure}

The branching ratios between the two transitions
$\sigma$(DC$\to$429)/$\sigma$(DC$\to$0) have been measured at E$_{\alpha}$ = 400, 250 and
220 keV and are  0.417 $\pm$0.020,
0.415$\pm$0.029 and 0.38$\pm$0.03, respectively. In Figure \ref{branching} the present data are compared to previous experimental results \cite{parker-exp,nagatani,osborne,krawinkel} and theoretical calculations \cite{liu,kajino1}.
Although our data improve the experimental precision at low energy, they are still compatible with both theoretical predictions.

\section{Data analysis of the prompt $\gamma$ experiment}

For an extended gas target, the number of detected photons
N$_{\gamma}$ is given by:

\begin{equation}
N_{\gamma}=N_p\int_{0}^{L}\sigma(E(z))\eta(z)\rho_{beam}(z)dz
\label{analysis}
\end{equation}

where N$_p$ is the number of the accelerated $\alpha$-particles
obtained from the calorimeter beam power measurement,
$\rho_{beam}$ is the effective target density that takes into
account the measured pressure, temperature profile and the beam heating
effect \cite{marta}, $\eta$(z) is the the detection efficiency and $\sigma$ is
the reaction cross section. The length L=80.2 cm is the distance between the first pumping stage and the calorimeter:
according to MC simulations this region corresponds to the gas target zone where 99.9\% of the detected fusion reactions take place.

Since the cross section is expected to be a smooth function at
low energies \cite{descouvemont,kajino}, an effective cross section
$\sigma_{eff}$ is introduced as the average cross section
over the interaction energies:

\begin{equation}\label{eq:sigma}
\sigma_{eff}=\frac{N_{\gamma}}{N_p\int_{0}^{L}\eta(z)\rho_{beam}(z)dz}
\end{equation}

From the definition of the S-factor \cite{rolfs}:

\begin{equation}
S(E)=\frac{\sigma(E)}{E}e^{-2\pi\eta(E)}
\label{eq:eq1}
\end{equation}

and equation \ref{eq:sigma}, one obtains the S(E$_{eff}$) factor, provided that an effective
interaction energy E$_{eff}$ is introduced \cite{albe}.

The effective energy E$_{eff}$ is defined by the relation \cite{albe}:

\begin{equation}\label{eq:e_eff}
\sigma(E_{eff})=\sigma_{eff}
\end{equation}

By inverting equation \ref{eq:e_eff}  one can obtain E$_{eff}$ from
$\sigma^{-1}$($\sigma_{eff})$. In our experimental conditions (i.e. gas pressure and beam energy), the target thickness $\Delta$E was around 10 keV, corresponding to 6.2$\times$10$^{17}$  At/cm$^2$.
Therefore, since theoretical models \cite{descouvemont,kajino} indicate a negligible S(E) energy dependence  inside $\Delta$E at these energies,
a constant S factor could be
considered in equation \ref{eq:eq1}, and the effective energy was
obtained from equation \ref{eq:e_eff} that reduces to:

\begin{equation}\label{eq:def}
\frac{e^{-2\pi\eta(E_{eff})}}{E_{eff}} =
\frac{\int_{0}^{L}\frac{e^{-2\pi\eta(E(z))}}{E(z)}\eta(z)\rho_{beam}(z)dz}{\int_{0}^{L}\eta(z)\rho_{beam}(z)dz}
\end{equation}

The uncertainty on the effective energy calculation is coming from the uncertainty on the beam energy (absolute calibration \cite{acceleratore}) and from the error on the energy lost by the beam inside the target \cite{ziegler}.


\section{Comparison between activation and prompt results}

The $^3$He($\alpha$,$\gamma$)$^7$Be data taking lasted several months. In a first phase only activation
measurements were performed: these results have been reported in
\cite{bemmerer,giuri}. Thereafter, a second phase (here detailed)
started aimed at studying the reaction using both
activation and prompt $\gamma$ method at the same time. Since the irradiation of
the samples used for off-line $^7Be$ counting were
simultaneously performed to the $\gamma$ radiation detection, some
systematic uncertainties are common to both methods and were not
considered in the comparison between the S factors obtained with
the two techniques. In table \ref{table1} the sources of systematic uncertainty affecting both methods
and their contribution to the final uncertainty on
the S-factor, are listed. 
All the LUNA results \cite{bemmerer,giuri,confortola} are collected in Table \ref{table2}.
The activation S-factor values obtained in the two phases of the
experiment at about the same beam energy are compatible (Table \ref{table2}).

\begin{table*}[here!]
\begin{center}
\begin{tabular}{l c c }
\hline
Source  &  Prompt   &   Activation \\
 \hline
Beam Intensity    &  1.5\%   & 1.5\%  \\
$^3$He Target Density  & 1.5\%   & 1.5\%\\
Effective Energy   & 0.5-1.1\%  & 0.5-1.1\% \\
Angular Distribution & 2.5\%  &     \\
Detection Efficiency & 1.5\%  &     \\
$^7$Be counting efficiency  &   & 1.8\% \\
Incomplete $^7$Be collection &   & 0.5\% \\
$^7$Be Backscattering  &    &  0.5\%\\
$^7$Be Distribution in catcher &  &0.4\% \\
478 keV $\gamma$-ray branching &    &  0.4\% \\
$^7$Be Half life    &  & 0.1\%\\
Parasitic $^7$Be production &  & 0.1\%\\
\hline

Total   &  (3.6-3.9)\%  & (3.0-3.2)\% \\
\hline

\end{tabular}
\caption{Systematic uncertainties and their contribution to the
S-factor error for the prompt and activation experiments.} \label{table1}
\end{center}
\end{table*}

\begin{landscape}
\begin{table*}[here!]
\begin{center}
\begin{tabular}{p{1cm} c c  l c c c c c c  }

E$_{eff}$ & Method  & Charge &   Peak &  Gross & Background & $\sigma$(E$_{eff}$)& S(E$_{eff}$) & $\Delta$S stat. & $\Delta$S syst. \\
(keV)     &         & (C)    &        &   Counts  &  Counts   &(nbarn) & (keVb)    &  (keVb)    & (keVb)   \\
\hline

170.1  &p    &  112.7 &   DC$\to$ 0 & 6780     & 89       & 7.23 $\pm$0.26 &  \raisebox{-1.5ex}{0.510} &  \raisebox{-1.5ex}{0.008} & \raisebox{-1.5ex}{0.019}  \\
&p        &          &  DC$\to$ 429 & 3500  & 666 & 3.02 $\pm$ 0.12 \\
169.5 & a&   112.9 &    478$\to$ 0 & 8666 &579 & 10.0 $\pm$0.35  & 0.507 & 0.010 & 0.015 \\

168.9 & a$^{\star}$  & 62.5          &     478$\to$ 0 & 7295 & 1161  &9.35 $\pm$0.19  & 0.482 & 0.02  &0.03  \\

147.7 & a$^{\star}$ & 203.1          &     478$\to$ 0 & 10551  & 1033& 4.61 $\pm$ 0.07 & 0.499  & 0.017  & 0.03 \\

126.5 & a$^{\star}$ & 215.7         &       478$\to$ 0 & 2866   &  95 & 1.87 $\pm$0.04 & 0.514 & 0.02  & 0.03 \\

106.1& p &  406.93 &    DC$\to$ 0 & 1516  & 67 & 0.415 $\pm$ 0.018 & \raisebox{-1.5ex}{0.518} & \raisebox{-1.5ex}{0.014} & \raisebox{-1.5ex}{0.020} \\
      & p &        &      DC$\to$ 429 & 745  & 142 & 0.173 $\pm$ 0.010 \\
105.7 & a&   413.6 &  478$\to$ 0 & 3764 &  1214& 0.546 $\pm$0.024 & 0.493 & 0.015 & 0.015 \\

93.3& p &    636.73 &   DC$\to$ 0 & 988 & 53 & 0.171 $\pm$ 0.008& \raisebox{-1.5ex}{0.527} & \raisebox{-1.5ex}{0.018} & \raisebox{-1.5ex}{0.021} \\
      & p  &       &  DC$\to$ 429 & 479  & 135 &0.065 $\pm$ 0.005 \\
92.9 &   a & 725.8 &  478$\to$ 0 &  5123&2473 &0.232  $\pm$ 0.01& 0.534 & 0.016 & 0.017 \\

\end{tabular}
\caption{Summary of the LUNA prompt (p) and activation (a) data: the symbol $\star$ indicates runs in which only activation data have been collected  \cite{bemmerer,giuri}.} \label{table2}
\end{center}
\end{table*}
\end{landscape}

\section{Comparison with other experiments}

In the last forty years the $^3$He($\alpha$,$\gamma$)$^7$Be
reaction has been extensively studied using both the activation
and prompt $\gamma$ detection method. An overall analysis
\cite{adelberger} showed an average discrepancy between S(0)
results obtained from the two methods of around 9\%. Starting
with the precise activation measurement in 2004 \cite{nara}, a
second generation
 of experiments has started with the aim of studying the
  $^3$He($\alpha$,$\gamma$)$^7$Be reaction with high accuracy.
Later, LUNA has
measured the reaction in two different experiments: first an activation
measurement with an accuracy of 3\% has been reached \cite{bemmerer,giuri}, and
subsequently the simultaneous activation and prompt measurement presented here, has been performed obtaining an average accuracy of 4\%.
 Most recently, a new
simultaneous activation and prompt measurement has been carried
out  \cite{brown}, that extends over a larger energy range going
from a minimum energy of E$_{cm}$=330 keV to a maximum energy of
E$_{cm}$=1230 keV. The data were measured with an accuracy of the
order of 3\% \cite{brown}.

\begin{figure}
    \includegraphics[width=\textwidth]{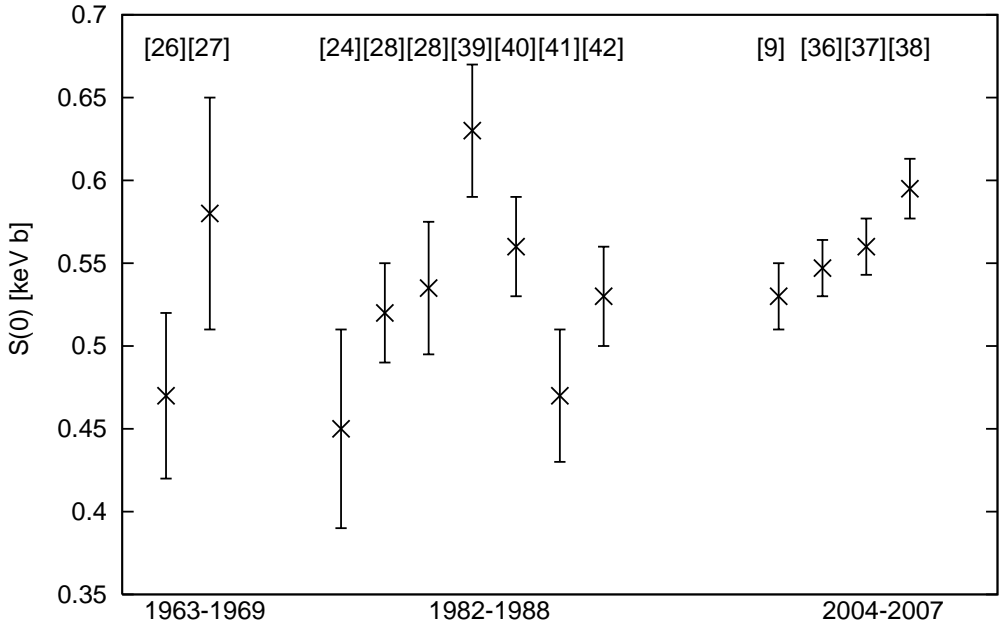}
    \caption{$^3$He($\alpha$,$\gamma$)$^7$Be S(0) values obtained from different experiments as a function of time.Both activation and prompt $\gamma$ experiments are considered.}
    \label{fig3}
\end{figure}

 In figure \ref{fig3} the time-trend of the S(0) values obtained from
different experiments is shown.
A clear evidence for the increase in the accuracy of the obtained S(0) is visible in the second generation experiments due to a better control on the
systematic effects which namely, could be the origin of the discrepancy between prompt and activation data claimed in the past \cite{adelberger}.
 Therefore we decided to consider only the data from the three most recent experiments \cite{nara,bemmerer,giuri,brown,confortola}.
Following the approach from \cite{brown}
we fitted the data of the different experiments using the
same theoretical curves. We used the resonating-group calculation
curve of Kajino et al  \cite{kajino} and the R-matrix fit of
Descouvemont et al. \cite{descouvemont}. Other theoretical trends for the S-factor are given in literature such as the one obtained with a cluster model calculation by Csoto and Langanke \cite{csoto}. This approach considers non external contributions to the cross section and therefore can not be normalized to the experimental data.

\begin{figure}
    \includegraphics[width=\textwidth]{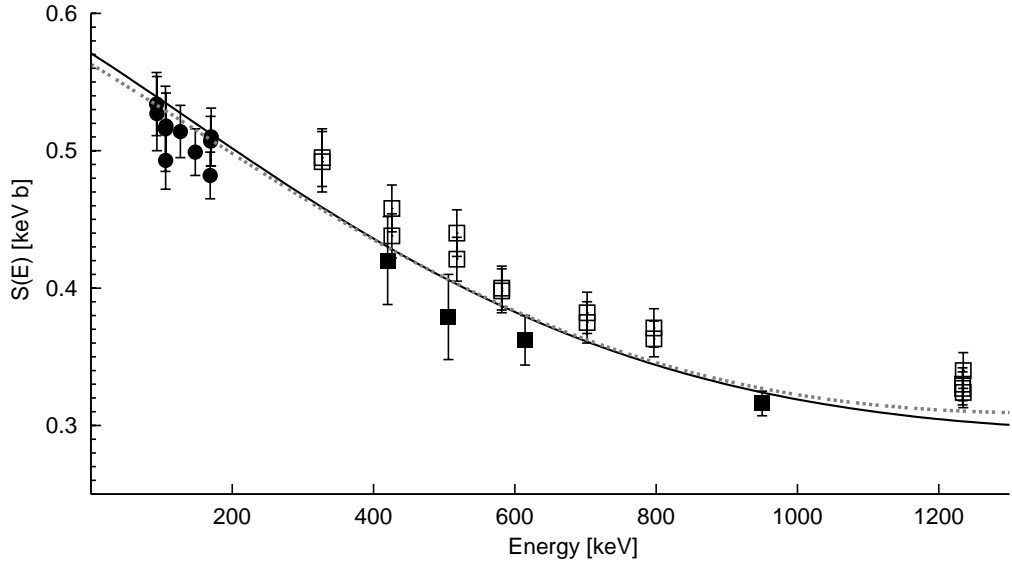}
    \caption{Astrophysical S factor for the $^3$He($\alpha$,$\gamma$)$^7$Be reaction obtained from the most recent experiments.
The filled squares are the data from  Singh et al. \cite{nara}, the filled circles are all the data from the LUNA experiment \cite{bemmerer,giuri,confortola}
 and the open squares are all the data from Brown et al. \cite{brown}. The solid and the dotted curves are the theoretical curves by \cite{descouvemont} and \cite{kajino} respectively, obtained considering the weighted average of S(0) of the different experiments as explained in the text and shown in Table \ref{tabella3}.
.}
    \label{S(E)}
\end{figure}

\begin{table*}[here!]
\begin{center}

\begin{tabular}{l c c c  }
\hline \hline
    &   S(0) keV barn       & S(0) keV barn     \\
    &   Kajino et al.\cite{kajino} & Descouvemont et al.\cite{descouvemont} \\
\hline \hline
LUNA activation data \cite{bemmerer,giuri,confortola} &  0.548$\pm$0.017 & 0.550$\pm$0.017 \\
LUNA prompt data \cite{confortola} & 0.561$\pm$0.021 & 0.564$\pm$0.021 \\
Singh et al. activation data \cite{nara} & 0.541$\pm$0.02 &0.551$\pm$0.02  \\
Brown et al. activation data \cite{brown}& 0.595$\pm$0.018 & 0.609$\pm$0.019 \\
Brown et al. prompt data \cite{brown}& 0.596$\pm$0.021 & 0.610$\pm$0.022 \\
\hline
S(0)$_P$ & 0.579$\pm$0.018 & 0.586$\pm$0.023\\
S(0)$_A$ & 0.562$\pm$0.017 & 0.569$\pm$0.019\\
\hline \hline
S(0)-TOT &   0.563$\pm$0.016   & 0.571$\pm$0.019        \\

\end{tabular}

\caption{S(0) values obtained rescaling the Kajino et al.
\cite{kajino} and the Descouvemont et al \cite{descouvemont}
theoretical curves. S(0)$_A$ value is obtained
from a weighted average of the S(0) values from the activation measurements
\cite{bemmerer,giuri,confortola,nara,brown} while S(0)$_P$ is the weighted average of the S(0) values from the prompt measurements
\cite{confortola,brown}.} \label{tabella3}
\end{center}
\end{table*}

The two theoretical
curves \cite{kajino,descouvemont} were re-scaled to the data of the different experiments
and the obtained  S(0) values are presented in table \ref{tabella3}.
S(0)$_A$ value is obtained
from a weighted average of the S(0) values from the activation measurements
\cite{bemmerer,giuri,confortola,nara,brown} while S(0)$_P$ is the weighted average of the S(0) values from the prompt measurements
\cite{confortola,brown}.
 The average discrepancy between prompt
and activation results $\Delta$S(0)=(S(0)$_A$-S(0)$_P$)/((S(0)$_A$+S(0)$_P$)/2))
is $\Delta$S(0)= -0.030$\pm$0.04 considering the Kajino et al. theoretical curve and is $\Delta$S(0)= -0.029$\pm$0.05 considering the Descouvemont et al. R-matrix fit.  This global result confirms
that no discrepancy is actually observable between results obtained from the two techniques and excludes significant non-radiative contributions to the reaction cross section.
Finally, we obtain a total S(0) = 0.563$\pm$ 0.016 keVb and S(0)= 0.571$\pm$0.019 keVb adopting the curves from \cite{kajino} and from \cite{descouvemont} respectively (see Table \ref{tabella3} and Figure \ref{S(E)}). These values are obtained from a weighted average of the S(0) value of \cite{nara},
of the S(0) value of LUNA (activation and prompt results combined \cite{bemmerer,giuri,confortola})
 and from the S(0) value of \cite{brown} (activation and prompt results combined).
The final errors on S(0)$_P$, S(0)$_A$ and S(0)-TOT are larger than the errors obtained from a simple weighted average. Since the scatter of the points about the mean is larger than expected based on the quoted errors, we have followed the method described in \cite{kajino,descouvemont,pdg} consisting in increasing the uncertainties on the single data so as to make the value of $\chi^2$ per degree of freedom equal to 1.0.

\section{Conclusions}

From an overall analysis of the results of the last generation experiments on  $^3$He($\alpha$,$\gamma$)$^7$Be, no discrepancy emerged
between prompt and activation data. Furthermore, a total (statistical and systematical) accuracy of about 3\% for
 the S(0) value was obtained: S(0)=0.567$\pm$0.018$\pm$0.004 keV b where the last term sizes the indetermination on the theoretical model adopted for the extrapolation to zero energy.
However, preliminary results recently obtained with the recoil separator technique between 1 and 3 MeV \cite{erna}, show a different S-factor energy dependence. Therefore, further improvements could come from new experiments exploring, with the same setup, the entire energy range from 0.1  to few MeV.

The present result lowers significantly the uncertainty coming from the {$^3$He($\alpha$,$\gamma$)$^7$Be nuclear reaction on the $^8$B and $^7$Be neutrino flux.
As described in a recent paper by Haxton and Serenelli \cite{serenelli} the solar interior metallicity can be obtained by measuring the solar CN neutrino flux. The latter can be related to the measured and predicted $^8$B neutrino flux, the predicted CN neutrino flux and the C and N abundances in the solar interior (eq. 9 and 13 in \cite{serenelli}). In the near future it should be possible to  measure the CN neutrino flux with experiments as Borexino \cite{borexino} and the upgraded SNO experiment \cite{chen}.
 Thanks to the low uncertainties now achieved on the measured rate of the key reactions $^3$He($\alpha$,$\gamma$)$^7$Be and $^{14}$N(p,$\gamma$)$^{15}$O and on the precise measurement of the $^8$B neutrino flux, Borexino and SNO could determine the C and N abundances in the radiative solar core. A comparison of the Sun's deep interior and surface
composition could be done, testing a key assumption of the standard solar model:
a homogeneous zero-age Sun. It would also provide a cross-check on recent photospheric
abundance determinations that have altered the once excellent agreement between the
SSM and helioseismology \cite{asplund}.

\section{Aknowledgements}
This work was supported by INFN and in part by the European Union
(TARI RII3-CT-2004-506222) and OTKA (K68801 and T49245), by ILIAS integrating activity
(Contract No.RII3-CT-2004-506222) as part of the EU FP6 programme and BMBF (05Cl1PC1-1).




\end{document}